\documentclass[journal=jpcafh,manuscript=article]{achemso}

\usepackage[version=4]{mhchem} 
\usepackage{fancyvrb}
\usepackage{booktabs}
\usepackage{graphicx}
\usepackage{amsmath}

\author{Michael McCarthy}
\affiliation{Center for Astrophysics $\vert$ Harvard \& Smithsonian, 60 Garden Street, Cambridge MA 02138, USA}
\author{Kin Long Kelvin Lee}
\affiliation{Center for Astrophysics $\vert$ Harvard \& Smithsonian, 60 Garden Street, Cambridge MA 02138, USA}
\email{kin_long_kelvin.lee@cfa.harvard.edu}
\title{Molecule Identification with Rotational Spectroscopy and Probabilistic Deep Learning}

\abbreviations{IR,NMR,UV}

\begin{document}

\begin{tocentry}
\includegraphics{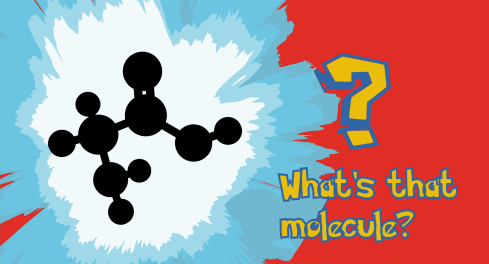}





\end{tocentry}

\begin{abstract}

A proof-of-concept framework for identifying molecules of unknown elemental composition and structure using  experimental rotational data and probabilistic deep learning is presented. Using a minimal set of input data determined experimentally, we describe four neural network architectures that yield information to assist in the identification of an unknown molecule. The first architecture translates spectroscopic parameters into Coulomb matrix eigenspectra, as a method of recovering chemical and structural information encoded in the rotational spectrum. The eigenspectrum is subsequently used by three deep learning networks to constrain the range of stoichiometries, generate SMILES strings, and predict the most likely functional groups present in the molecule. In each model, we utilize dropout layers as an approximation to Bayesian sampling, which subsequently generates probabilistic predictions from otherwise deterministic models. These models are trained on a modestly sized theoretical dataset comprising ${\sim}$83,000 unique organic molecules (between 18 and 180\,amu) optimized at the $\omega$B97X-D/6-31+G(d) level of theory where the theoretical uncertainty of the spectoscopic constants are well understood and used to further augment training. Since chemical and structural properties depend highly on molecular composition, we divided the dataset into four groups corresponding to pure hydrocarbons, oxygen-bearing, nitrogen-bearing, and both oxygen- and nitrogen-bearing species, training each type of network with one of these categories thus creating ``experts'' within each domain of molecules. We demonstrate how these models can then be used for practical inference on four molecules, and discuss both the strengths and shortcomings of our approach, and the future directions these architectures can take.

\end{abstract}

\section{Introduction}

The ability to determine the elemental composition and three-dimensional structure of an unknown molecule is highly relevant in nearly all fields of chemistry. 
Microwave spectroscopy has many favorable attributes in this regard because its spectral resolution is intrinsically very high and because rotational transition frequencies sensitively depend on the geometry of the molecule. For these reasons, it has been used with good success in characterizing mixtures containing both familiar and unknown species. With the development of broadband chirped-pulse methods,\cite{brown_broadband_2008, park_perspective_2016, wehres_100_2017, finneran_direct_2013} microwave instruments can routinely sample an octave or more of frequency bandwidth while simultaneously achieving parts per million resolution at low pressure. Under these conditions it is relative straightforward to distinguish between two molecules with very similar structure, and, as a consequence, gas mixtures containing in excess of 100 different compounds have analyzed\cite{lee_study_2019} using highly automated experimental techniques and methodologies.\cite{crabtree_microwave_2016,drumel_automated_2016,zaleski_automated_2018,seifert_autofit_2015,western_pgopher:_2017,riffe_rotational_2019}

As the throughput of microwave spectrometers continues to increase, data analysis rather than acquisition has become the primary obstacle to translating spectral information into chemical knowledge. From a scientific and analytical standpoint, the ability to analyze complex mixtures in real-time would substantially improve the rate of discovery, especially with respect to identifying unknown species in heavily congested spectra. Typically, deconvolution of a mixture is performed by spectrally separating rotational transitions of an individual chemical constituent, isotopic species, or excited state, based on the small number of spectroscopic constants that are needed to reproduce its rotational spectrum.  Most prominent among these are the three rotational constants ($A(BC)$) which are inversely proportional to the principal moments of inertia, and thus encode the distribution of mass in three-dimensional space. Indeed, these parameters are widely used in experimental molecular structure determination through a variety of methods.\cite{bohn_second_2016,demaison_equilibrium_2011}

The conventional process of identifying an unknown molecule by microwave spectroscopy involves comparing the magnitudes of the three experimental rotational constants with those predicted by electronic structure calculations for a series of candidate molecules, most of which are selected on the basis of chemical intuition. Intuition, in this case, requires consideration of the likely elemental composition, starting precursors, experimental conditions, and well-characterized molecules with similar rotational constants. For small systems where there are relatively few possible combinations, chemical intuition will often narrow down candidates quickly, and the number of electronic structure calculations required is small. For heavier and larger molecules, the number of possible structures grows rapidly with respect to composition and structural diversity. In truly unknown analytical mixtures where information is limited---such as those encountered in electrical discharge experiments\cite{lee_study_2019} and astronomical observations---the combinatorics become intractable and chemical intuition is both highly inefficient and incomplete with respect to capturing the full range of possible outcomes.

Machine learning (ML) is an attractive tool to assist in the identification of newly discovering molecules. At a high level, ML methodologies learn a set of parameters, $\theta$, that are then used to estimate some property $y$ that can help assist the identification of a molecule based on its spectroscopic data $x$. Here, $y$ ideally represents a three-dimensional molecular structure, which---using rare-isotopic spectroscopy---can be directly confirmed experimentally based on the expected shifts in rotational constants. Other discerning factors that can be substituted for $y$ include possible elemental composition, presence of functional groups, and the number of non-hydrogenic atoms.

To identify molecules solely from available spectroscopic information, we require a ML methodology that can satisfy two criteria: first, it must encapsulate all of the possible structural and chemical space for a given set of $A(BC)$, that is molecules with different compositions and structures can have the similar rotational constants; second, the method must provide some estimate of uncertainty. The first criteria ensures that the method can break the partial degeneracy of $A(BC)$ where the composition is not necessarily known and may represent entirely different molecules and structures. The second criteria is necessary to infer possible carriers; it is impossible to deterministically know the exact carrier simply from $A(BC)$, and instead it must be taken from a distribution of possible candidates.

Probabilistic neural networks\cite{goodfellow_deep_2016} are an extremely felicitous class of ML techniques that provide solutions relevant to both criteria. Built on top of conventional deep learning models which learn from a training set of data and provide the maximum likelihood estimate, probabilistic approaches ultimately yield a distribution of predictions weighted and their associated likelihoods. With a sufficiently large and diverse data set of information, a probabilistic neural network model can be trained to transform spectroscopic parameters $x$ into discerning information $y$. Formally, the problem of molecular identification then becomes estimating the conditional likelihood $p(y \vert x, \theta)$---the likelihood an unknown molecule with parameters ($x$) can be identified with information $y$ based on learned parameters $\theta$.

As a proof-of-concept for the usefulness of probabilistic deep learning in molecule identification, we combine ensembles of relatively simple neural network architectures with computationally cost effective approximations to Bayesian sampling via dropout layers.\cite{gal_dropout_2015} Each model within the ensemble is trained on electronic structure calculations comprising a specific chemical composition (i.e. pure hydrocarbons, oxygen-bearing molecules) as a way to break the chemical/structural degeneracy of $A(BC)$, such that each respective model becomes an ``expert''. In essence, each model yields conditional predictions that correspond to a particular composition; for example, predicted $y$ for a given set of constants $x$, if the molecule was a pure hydrocarbon. The first model we consider translates spectroscopic parameters into Coulomb matrix eigenvalues as a way of decoding spectroscopy data into machine representations that encode molecular structure and chemical properties. The predicted eigenspectra are subsequently used by three independent models that predict the possible molecular formulae, functional groups present, and SMILES encoding for a given composition. The ability to determine composition and functionalization is not only useful for identification, but also deepens the connection with other analytical techniques such as mass spectrometry and infrared spectroscopy. The latter, in particular, we show can only be accessed through our new deep learning framework, and unlocks a new facet of rotational spectroscopy. The early sections of this paper will detail the expected results and performance of each model, and where applicable, comparison with a baseline ML model. In the last section, we discuss how the information from these models can be collectively interpreted in order to infer the identity of unknown molecules.

\section{Methodology}

\subsection{Molecule generation}

In order to train the deep learning models, we required a dataset of molecules that span a sufficiently large volume of structural and chemical space. Initial structures were generated via two mechanisms: by parsing SMILES published in the PubChem database and by systematic generation with the Open Molecule Generator (OMG)\cite{peironcely_omg:_2012}. For both cases, we systematically generated hundreds of formulae pertaining to simple organic species that constitute \ce{H_wC_xO_yN_z}, where $1\leq w \leq 18$, $1 \leq x \leq 8$, and $y, z \leq 3$ and with an even number of electrons, and where $w \geq x + y + z$. As the number of isomers grows combinatorially with the number of atoms, many formulae generate up to hundreds of thousands of possible SMILES codes -- to keep the number of quantum chemical calculations tractable, we truncate the largest lists and instead randomly sample up to 2,000 SMILES with uniform probability as a method of taking representative species for a given formula. Over the course of training we observed that the dataset underrepresented pure hydrocarbon species: subsequently, we bolstered the hydrocarbon set by generating isomers up to \ce{H20C10} using OMG.

Cartesian coordinates are generated from the SMILES codes using OpenBabel,\cite{oboyle_open_2011} which are subsequently refined using electronic structure calculations with Gaussian `16\cite{frisch_gaussian_2016} at the $\omega$B97X-D/6-31+G(d) level of theory and optimizing the geometry corresponding to the lowest singlet state. This method was chosen based on earlier benchmarking from which Bayesian uncertainties were obtained for several low-cost methods and basis sets, comparing the theoretical equilibrium rotational constants with vibrationally-averaged experimental values.\cite{lee_bayesian_2020} These results showed that $\omega$B97X-D/6-31+G(d) provided an excellent compromise between low theoretical uncertainty, good accuracy with respect to experimental constants, and low computational cost. Additionally, as we shall see later, the uncertainties are also used to augment our training sets.

\subsection{Data preprocessing}

Upon completion, the electronic structure calculation outputs are parsed, extracting relevant information such as the electronic energy, spectroscopic constants, harmonic frequencies, electric dipole moments, the corresponding canonical SMILES string using OpenBabel \cite{oboyle_open_2011}, and Cartesian coordinates of the molecule in the principal axis orientation. The results are then filtered, removing non-convergent structures, transition state structures and duplicate species by comparing the rotational constants and dipole moments. To facilitate the ensemble of models, we categorized the molecules in the dataset into four groups based on their composition: pure hydrocarbons (\ce{HC}), oxygen-bearing (\ce{HCO}), nitrogen-bearing (\ce{HCN}), and oxygen- and nitrogen-bearing (\ce{HCON}).

The optimized Cartesian coordinates are used to calculate the corresponding Coulomb matrix $M_{ij}$,\cite{rupp_fast_2012} defined by:

\begin{equation}
    M_{ij} = \begin{cases} 
    0.5Z^{2.4}_i & \text{for}~i = j\\
    \frac{Z_iZ_j}{\vert \mathbf{R}_i - \mathbf{R}_j} & \text{for}~i \neq j\\
    \end{cases}
\end{equation}

where $i,j$ are atom indices, $Z$ the atomic number, and $\mathbf{R}$ the coordinates of a given atom. The matrix maps the three-dimensional charge distribution of a molecule into a symmetric, two-dimensional projection of shape $n\times n$, where $n$ is the number of atoms. This machine representation of molecular structure is simultaneously unique in representation (apart from enantiomers) and encodes a significant amount of chemical information.

\begin{figure}
    \centering
    \includegraphics[width=0.6\textwidth]{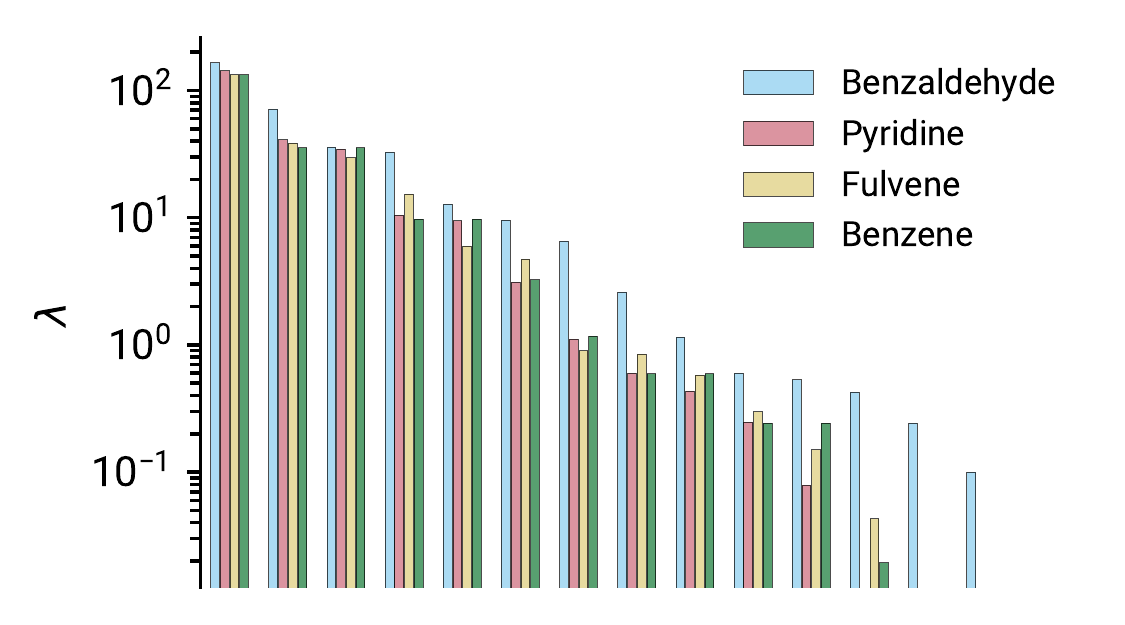}
    \caption{Comparison of eigenspectra for four structurally and chemically similar species. The spectrum is truncated to the first fifteen non-zero elements.}
    \label{fig:ringspectra}
\end{figure}

Because the experimental data typically consists of only up to eight parameters, there is a need to reduce the dimensionality of our molecular representation: a set of rotational constants is unlikely to effectively sample all possible Coulomb matrix configurations; instead, we choose to use the eigenvalues $\lambda = [\lambda_i \ldots \lambda_n]$ of the Coulomb matrix. While the absolute positions of atoms are lost, the maximum value and the decay in magnitude of the eigenspectrum reflects the type of atoms present, as well as the general size of the molecule: smaller molecules display ``shorter'' eigenspectra compared to larger species, and molecules that contain more non-hydrogen atoms demonstrate slower decaying eigenspectra with larger magnitude eigenvalues. Despite a reduction in dimensionality, Figure \ref{fig:ringspectra} shows eigenspectra can still readily differentiate between even similar molecules---fulvene is a higher-energy isomer of benzene, pyridine is isoelectronic with benzene, and benzaldehyde is a functionalized derivative of benzene. While the magnitude of the leading eigenvalues are similar, they are differentiable particularly towards the tail end of the eigenspectra; for example, the eigenspectrum continues for benzaldehyde, whereas the spectrum of pyridine truncates earlier.

\begin{figure}
    \centering
    \includegraphics[width=0.6\textwidth]{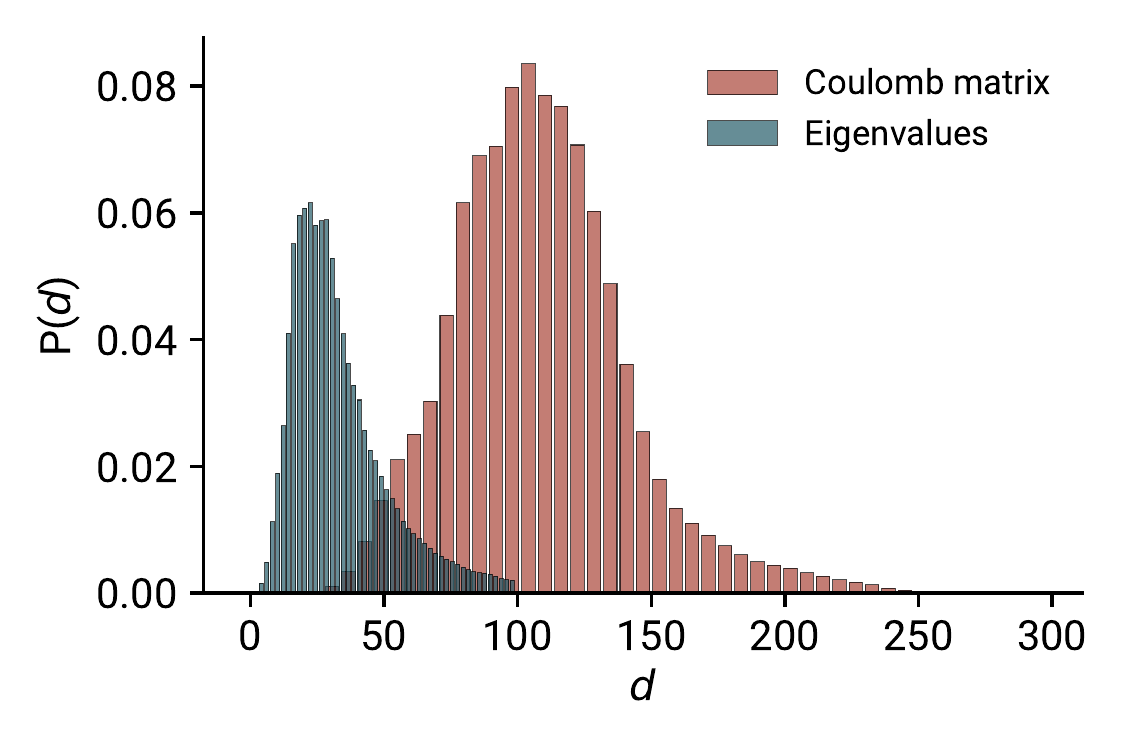}
    \caption{Pairwise similarities of molecules within the dataset measured by Euclidean ($l^2$) distance. }
    \label{fig:distances}
\end{figure}

Figure \ref{fig:distances} compares all of the pair-wise euclidean ($l^2$) distances between molecules within the dataset. In both the Coulomb matrix and the reduced eigenspectrum representation, the distributions peak far from zero and thus expected to be readily differentiable by machine learning models. The distribution of distances is similar to those seen in other large organic molecule datasets such as QM9 \cite{ramakrishnan_quantum_2014, ruddigkeit_enumeration_2012}.

In order for the model to process the formula, SMILES strings, and functional groups we converted these labels into corresponding vector representations. For each molecule, the chemical formula is encoded into a length four vector, where the index corresponds to atom symbol and value the number of the corresponding atom. With respect to SMILES, we used one-hot encoding similar to those demonstrated in several studies \cite{gupta_generative_2018,hirohara_convolutional_2018} -- each SMILES string is encoded in a two-dimensional matrix where rows correspond with character, and each column index represents one of the 29 SMILES symbols within our dataset corpus. The first column index is reserved for blank spaces, which are used to pad shorter SMILES strings up to 100 characters. The resulting SMILES arrays have a shape of $100\times30$. Finally, the functional group labels are generated based on the OpenBabel canonical SMILES strings by performing functional group substructure searches with the SMARTS language implemented in  RDKit.\cite{baltruschat_rdkit_2019} The functional groups within each molecule is subsequently represented as a multilabel, ``multihot'' encoding. A full table summarizing the encodings can be found in the Supporting Information.

\subsection{Neural network details}

The neural network models described in this work were implemented with PyTorch \cite{paszke_pytorch_2019}, with training performed on an Nvidia V100 GPU on the ``Hydra'' computing cluster at the Smithsonian Institution. In all cases, training was performed using the Adam optimizer \cite{kingma_adam:_2017} with decoupled weight decay. Training is performed on a 80:20 split, where 20\% of the dataset is held for validation between training epochs. At the end of each epoch, the training set is shuffled, such that each minibatch is different between passes. 

To improve generalization and uncertainty in each of the models, we have also adopted two augmentation strategies. First, it became apparent during development that the dataset was extremely imbalanced, despite the random and unbiased sampling approach we adopted during its creation. This was a direct consequence of the number of possible isomers for certain functional groups over others: for example, there are many more ways to form an amine (i.e. primary, secondary, tertiary) than a nitrile, which was frequently observed during inference. This is commonly encountered in multilabel classification, where there are insufficient examples of underrepresented labels for models to learn from and subsequently predict. To alleviate this, we duplicated species with functional groups that have fewer than 5,000 samples, and by adding Gaussian noise to the rotational constants and dipole moments we create ``new'' synthetic samples to balance the dataset. 

The second augmentation strategy was to apply data transformations between training epochs, which is done to mitigate overfitting and---particularly important in our application---to decrease model overconfidence. This method is commonly used in image-based applications, whereby adding Gaussian noise or random rotations improves the effective dataset size, and prevents overfitting. In our case, the rotational constants are augmented by the theoretical uncertainty associated with the electronic structure method used ($\omega$B97X-D/6-31+G(d)): the values of $A(BC)$ are scaled by a ratio $\delta$ sampled from a posterior likelihood $p(\delta)$ that represents the spread in discrepancy between the theoretical equilibrium rotational constants and the experimental vibrationally-averaged values.\cite{lee_bayesian_2020} In principle, this allows for model training to be performed on a ``vibrationally-averaged'' data which would otherwise be too costly to compute for the entire dataset. For the Coulomb matrix eigenspectra, we included Gaussian noise scaled by an exponential decay factor that preserves the tail seen in eigenspectra.

In all the architectures we explore in this work, each fully connected layer is paired with a dropout layer: for most cases, dropout layers act as a method of enforcing regularization during training by deactivating connections through each pass according to some probability $p$ \cite{hinton_improving_2012}. As an alternative purpose, \citet{gal_dropout_2015} has shown that during the prediction phase dropout layers can empirically approximate Bayesian sampling in Gaussian processes---providing $p$ is sufficiently large as to introduce enough stochasticity while maintaining accuracy. This approach emulates ensemble-based methods, whereby dropping different neurons with each forward pass effectively creates a sub-network. In our regression and recurrent models, these dropout units remain active during the prediction phase as a way to estimate model uncertainty with $p\approx0.3$ (i.e. each layer drops around 30\% of the units with each pass).

While dropout is a computationally efficient and simple way of determining uncertainty, this approach is known to underestimate model uncertainty.\cite{li_dropout_2017} Consequently, a single deep learning model with dropouts may not necessarily capture the full range of possible molecules based only on spectroscopic constants. As we shall see later, there are structural and chemical subtleties associated with molecules of varying compositions (e.g. oxygen-bearing vs. pure hydrocarbons) that force models to place varying importance on different parameters. To help alleviate this, we also employ an ensemble of networks---in general applications, this approach involves dividing the training data among multiple networks. As each network is exposed to a different dataset, the trained weights and biases differ, with the joint prediction having a smaller generalization error than a single network \cite{zhou_ensembling_2002, goodfellow_deep_2016}. In our application, each network is exposed to a specific composition of molecules that fall under the four categories mentioned previously, with the goal of preserving domain-specificity; that is, the same set of rotational constants can result from different chemical compositions, and need to be reflected in the model sampling. The premise is to learn and predict a given molecular property, if the unknown molecule were to contain a particular composition.

\begin{figure}
    \centering
    \includegraphics[width=\textwidth]{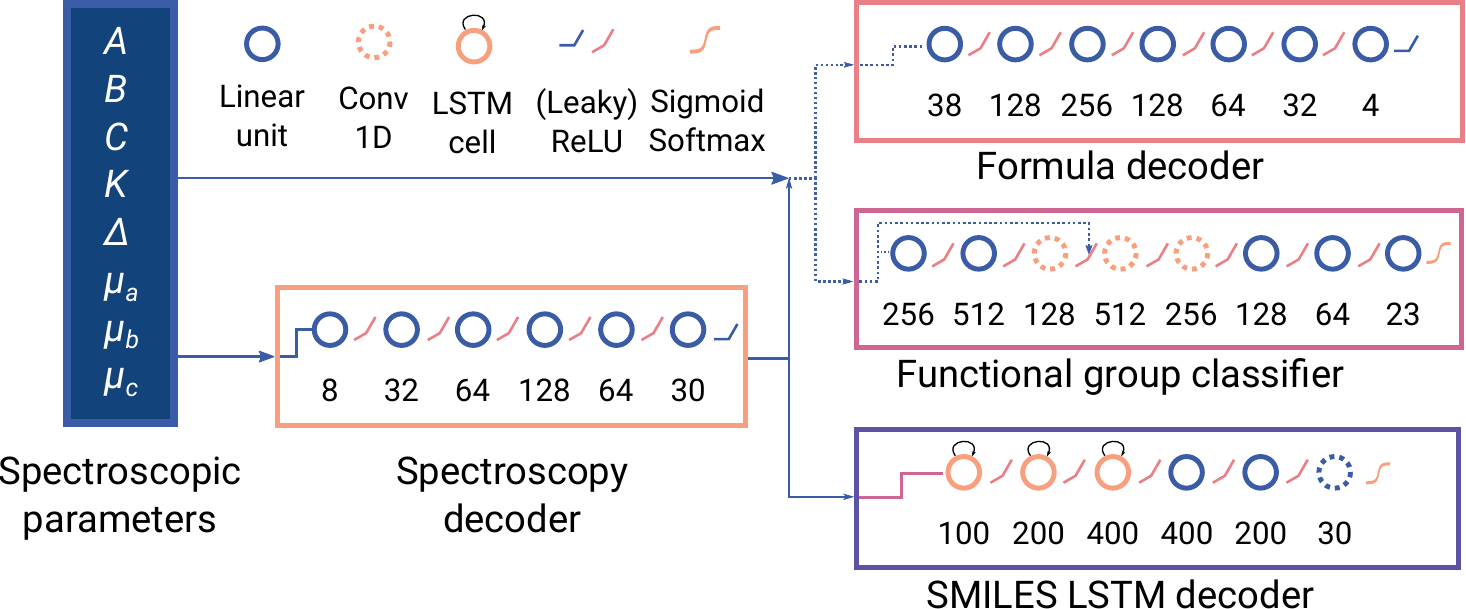}
    \caption{Graph depiction of the models considered in this work, with data flowing from left to right. Nodes represent each layer type, with the corresponding output size written below each node. Blue dotted lines represent the concatenated output of the spectroscopy decoder and the parameter inputs. The pink line within the SMILES LSTM decoder model corresponds to the timeshifted sequences of eigenvalues (see text).}
    \label{fig:architecture}
\end{figure}

Figure \ref{fig:architecture} shows the overall flow of data through the network models considered in this work. A user provides spectroscopic data that \textit{can} be experimentally derived, which is then used by the network to perform inference on the range of possible molecular formulae, generate viable SMILES codes, and predict the likelihood of functional groups present. In the case of the regression models, the architectures are relatively simple MLPs, up to seven layers deep and a maximum of 256 units wide, using LeakyReLU activation \cite{nair_rectified_2010,maas_rectier_2013} with a negative gradient of 0.3. The model training is performed by minimizing the mean absolute error between the model output and the regression targets (eigenspectra $\lambda$ and composition).

For the SMILES decoder, each LSTM cell used $\tanh$ and sigmoid functions for the cell and recurrent activations respectively. The output of the SMILES decoder corresponds to an array of shape $100\times30$, with each row corresponding to the likelihood distribution of a given SMILES character. The model is subsequently trained by minimizing the Kullback-Leibler (KL) divergence\cite{kullback_information_1978}:

\begin{equation}
    D_\mathrm{KL} = \sum_N \sum_M p(y \vert \lambda_m) \log \frac{p(y \vert \lambda_m)}{p_{\theta}(y \vert \lambda_m)}
\end{equation}

where $p_{\theta}(y \vert \lambda)$ represents the model softmax output, and $p(y \vert \lambda)$ the one-hot encoded SMILES encoding, for a given eigenspectrum $\lambda$. The loss is calculated by summation over $N$ minibatches comprising $M$ spectra. To help mitigate model overconfidence \cite{muller_when_2019}, we performed label smoothing on the SMILES encoding \cite{szegedy_rethinking_2015} whereby the one-hot encoded ground truth $p(y \vert \lambda)$ (which are Dirac delta functions) is smoothed by weighted uniform noise $\varepsilon u(k)$:

\begin{equation}
    p'(y \vert \lambda) = (1 - \varepsilon)\delta_{k,y} + \varepsilon u(k)
\end{equation}

where $k$ corresponds to the character label, uniform noise $u(k) = 1 / 30$, and weighting value $\varepsilon=0.1$. Consequently, the learning targets are no longer binary and forces the model to produce higher entropy/uncertainty predictions.

In the case of the functional groups, the task was to perform multilabel classification; training was performed by minimizing the binary cross-entropy loss. The architecture we propose here includes three 1D convolution layers, under the premise that the convolution kernels will learn characteristic relationships between the eigenspectrum and the spectroscopic parameters. Indeed, preliminary testing with simple MLP models (without convolution) performed significantly worse than the $k$-nearest neighbor baseline. In terms of activation functions, each convolution unit uses LeakyReLU ($\alpha=0.3$), whereas linear layers use parametric ReLU (PReLU) activations\cite{he_delving_2015} with the exception of the final output layer, which uses sigmoid activation. To characterize the classification performance, we computed the $F_1$ score\cite{manning_introduction_2008} across the full validation at the end of training:

\begin{equation}
    F_1 = \frac{2 \mathrm{P} \mathrm{R}}{ \mathrm{P} + \mathrm{R}}
    \label{eq:f1}
\end{equation}

where P and R are precision and recall scores; the former measures the number of times a correct label is applied out of all attempts, while the latter reports the ratio of correct labels predicted, out of all possible examples of a given label:

\begin{equation}
    \label{eq:pandr}
    \begin{split}
        \mathrm{P} & = \frac{T_\mathrm{p}}{N} \\
        \mathrm{R} & = \frac{T_\mathrm{p}}{T_\mathrm{p} + F_\mathrm{n}}
    \end{split}
\end{equation}

where $T_\mathrm{p}$, $F_\mathrm{n}$, and $N$ are the number of true positives, false negatives, and total number of samples respectively.\cite{manning_introduction_2008}

As these models represent a proof-of-concept, we have not extensively characterized or optimized hyperparameters nor architecture, with the exception of those encountered during training such as learning rate, minibatch size. The training parameters used are organized in Table \ref{tab:trainingparams}. In terms of the number of training epochs, each model was trained until the loss appears to have effectively converged, and there was no clear evidence for overfitting in neither the training/validation loss nor the prediction results. A large value of the weight decay ($\Lambda$) was used for each of the models as it drastically decreased model overconfidence---a known consequence of using dropouts to approximate Bayesian sampling.\cite{gal_theoretically_2016}

\begin{table}[]
    \centering
    \begin{tabular}{l r r r r r r}
        \toprule
        Model & $\alpha$ & $\Lambda$ & $N$ & Epochs & Loss & No. Parameters \\
        \midrule
        Spectroscopy decoder & $3\times10^{-3}$ & $10^{-1}$ & 100 & 80 & MAE & 20,896  \\
        Formula decoder & $5\times10^{-3}$& $2\times10^{-2}$ & 30 & 20 & MAE & 15,120 \\
        SMILES decoder & $10^{-3}$ & $10^{-1}$ & 500 & 30 & KL-Divergence & 579,588 \\
        Functional classifier & $1\times10^{-5}$ & $1\times10^{-1}$ & 300 & 50 & Cross-entropy & 668,443 \\
        \bottomrule
    \end{tabular}
    \caption{Summary of training parameters for each respective model. $\alpha$ and $\Lambda$ corresponds to the learning rate and weight decay respectively defined in the Adam optimizer\cite{loshchilov_decoupled_2019,kingma_adam:_2017}. $N$ refers to the minibatch size.}
    \label{tab:trainingparams}
\end{table}

\section{Results and discussion}

\subsection{Electronic structure calculations}

\begin{figure}
    \centering
    \includegraphics[width=\textwidth]{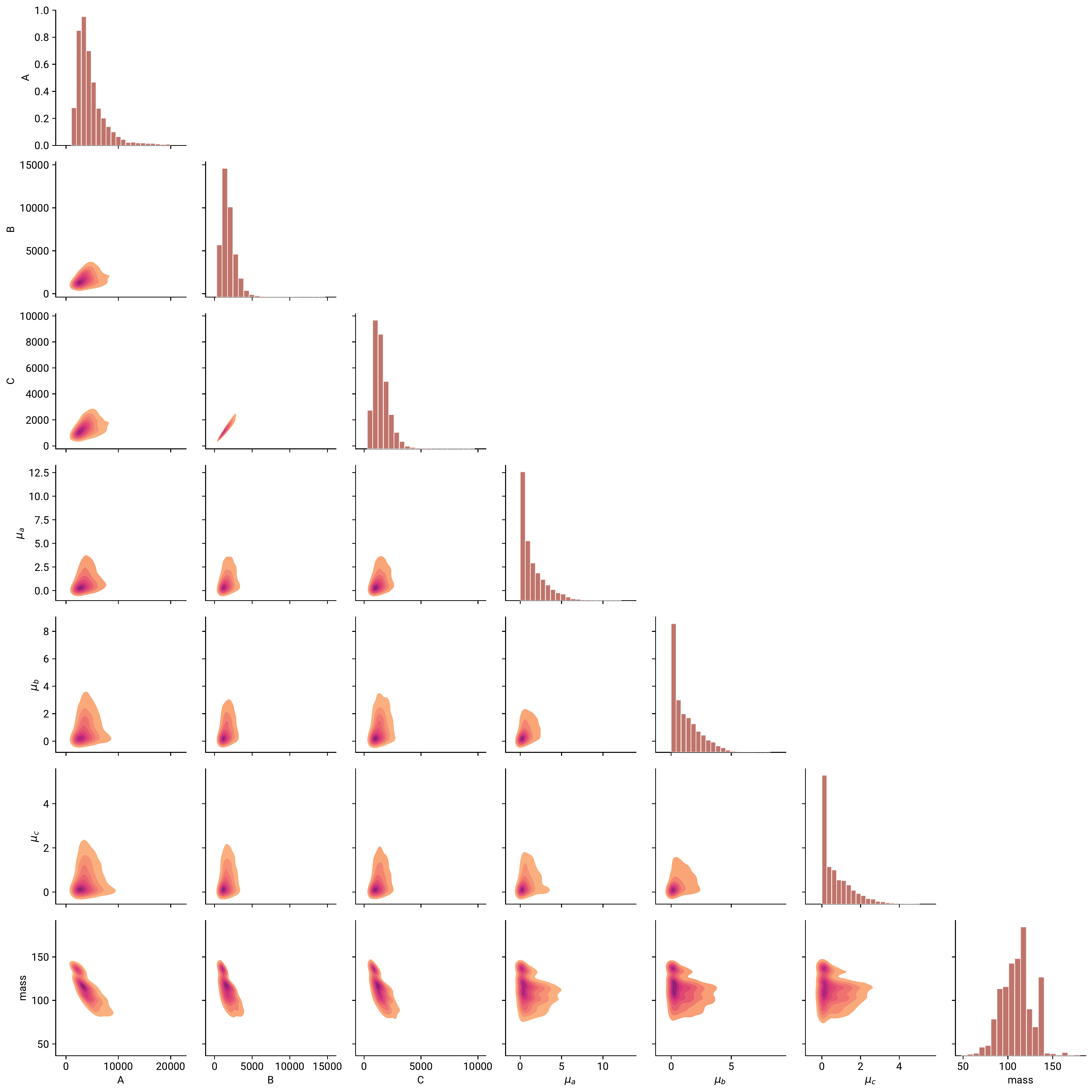}
    \caption{Visualization of the parameter space comprised by the dataset used for model training. Lighter species with rotational constants greater than 20,000 MHz are excluded in the visualization. Diagonal plots are histograms of features, while off-diagonal elements show density contours for pairs of features. The dipole moments correspond to their absolute values.}
    \label{fig:pairplot}
\end{figure}

Figure \ref{fig:pairplot} shows a correlation plot of the dataset parameters. With the exception of the rotational constants and molecular mass which are co-dependent, we see that all of the parameters are effectively uniformly distributed, and span a representative space along their respective dimensions. The rotational constants, particularly $B$ and $C$, decrease sharply with the molecular mass. The average species in our dataset is a near-prolate symmetric top ($\kappa < 0$, non-planar ($\Delta << 0$) with non-zero dipole moments along each axis and a mass of 108\,amu.

Table \ref{tab:datasetsummary} shows the summary statistics for the dataset, which provides another perspective to that seen in Figure \ref{fig:pairplot}. The mean and median (P50) are in qualitative agreement: most molecules in the dataset are near the prolate limit (i.e. $A>>B\approx C$) according to the asymmetry parameter. The average molecule possesses dipole moments along all three principal axes, on the order of a Debye for $\mu_a$ and $\mu_b$. Regarding the extremities, the lightest molecule in the dataset is \ce{CH2}, with the correspondingly largest rotational constants; the heaviest molecules considered (180\,amu) correspond to a formula \ce{H8C8O3N2}.

\begin{table}[]
    \centering
    \begin{tabular}{lrrrrrrr}
        \toprule
        Parameter & Mean & Std. dev. & Min. & P25 & P50 & P75 & Max. \\
        \midrule
        $A$ (MHz) &  5623.44 &  9949.31 &  820.76 &  2968.19 &  4141.94 &  6169.82 &  673708.06 \\
        $B$ (MHz) &  1851.40 &  1735.75 &  228.55 &  1136.62 &  1592.84 &  2259.91 &  337985.92 \\
        $C$ (MHz) &  1494.23 &  1199.04 &  225.28 &   951.87 &  1302.37 &  1823.59 &  213579.84 \\
        $\mu_a$ (D)  &     1.45 &     1.53 &    0.00 &     0.33 &     0.93 &     2.06 &      17.73 \\
        $\mu_b$ (D) &     1.17 &     1.18 &    0.00 &     0.25 &     0.81 &     1.74 &      11.81 \\
        $\mu_c$ (D) &     0.72 &     0.80 &    0.00 &     0.09 &     0.46 &     1.14 &       6.59 \\
        $\kappa$ &    -0.67 &     0.37 &   -1.00 &    -0.93 &    -0.81 &    -0.55 &       1.00 \\
        $\Delta$ (amu \AA$^2$) &   -61.47 &    49.17 & -388.87 &   -91.97 &   -48.27 &   -24.57 &       0.00 \\
        $M$ (amu) &   108.38 &    18.72 &   14.03 &    96.17 &   108.18 &   119.16 &     180.16 \\
        \bottomrule
    \end{tabular}
    \caption{Summary statistics of the parameters relevant to this study. $\kappa$, $\Delta$, and $M$ refer to the asymmetry parameter, the inertial defect, and the molecular mass respectively. P25, P50, and P75 correspond to the 25th, 50th, and 75th percentiles.}
    \label{tab:datasetsummary}
\end{table}

\subsection{Spectroscopy decoder}

\begin{figure}
    \centering
    \includegraphics[width=\textwidth]{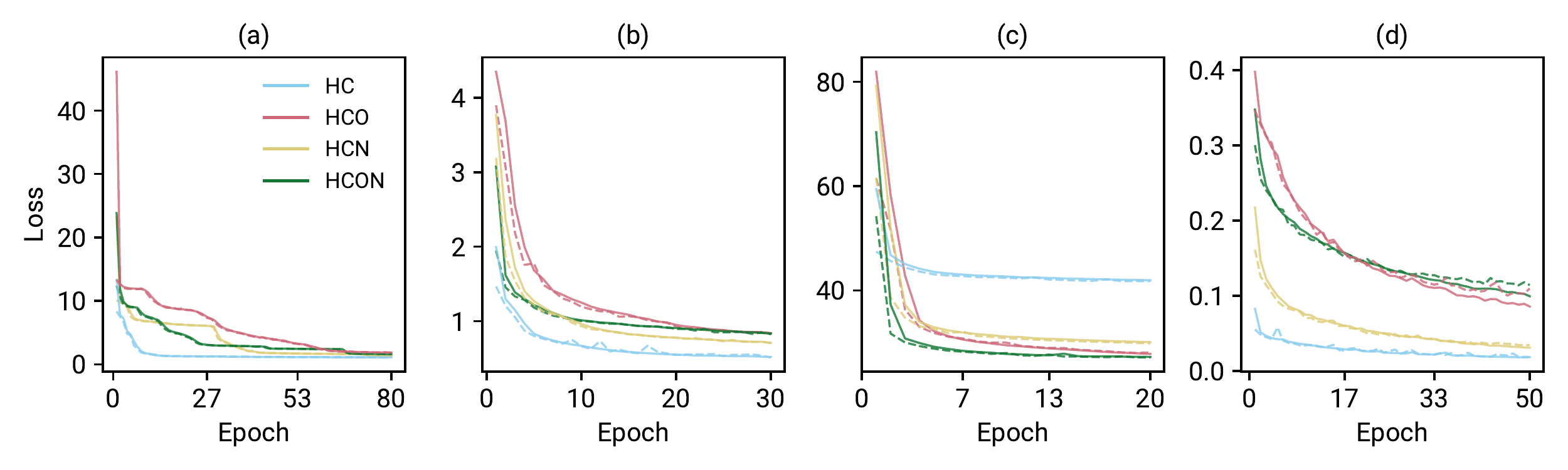}
    \caption{Epoch training (solid lines) and validation (dashed lines) loss for each of the models considered averaged across minibatches. Each color corresponds to a model composition: blue for pure hydrocarbons, red for oxygen-bearing, yellow for nitrogen-bearing, and green for oxygen- and nitrogen-bearing molecules. Panels (a) and (b) show the MAE loss, while (c) shows the KL-divergence and (d) shows the binary cross-entropy.}
    \label{fig:trainingerror}
\end{figure}

The first step in our approach involves taking experimental data as input, and encoding it as Coulomb matrix eigenspectra, which is responsible for translating spectroscopic constants into structural and chemical information. On a GV100, training over 80 epochs was completed for all four models in ${\sim}$30 minutes. Figure \ref{fig:trainingerror}(a) shows the training progress of the decoder model over 80 epochs, where the color traces represent ensemble sub-networks trained on a particular composition. The loss profiles appear turbulent, which reflects the difficulty in conditioning network parameters to map spectroscopic constants to the corresponding eigenvalues. Presumably, the learning rate would be a highly critical factor in the ultimate performance of this decoder model---currently, the final MAE are on the order of less than 1\% of the typical leading eigenvalues, which we believe provides sufficient accuracy for the subsequent decoding steps. The loss profiles suggest that the models are currently neither under- nor overfit, and thus could be readily extended in learning capacity.

\begin{figure}
    \centering
    \includegraphics[width=0.7\textwidth]{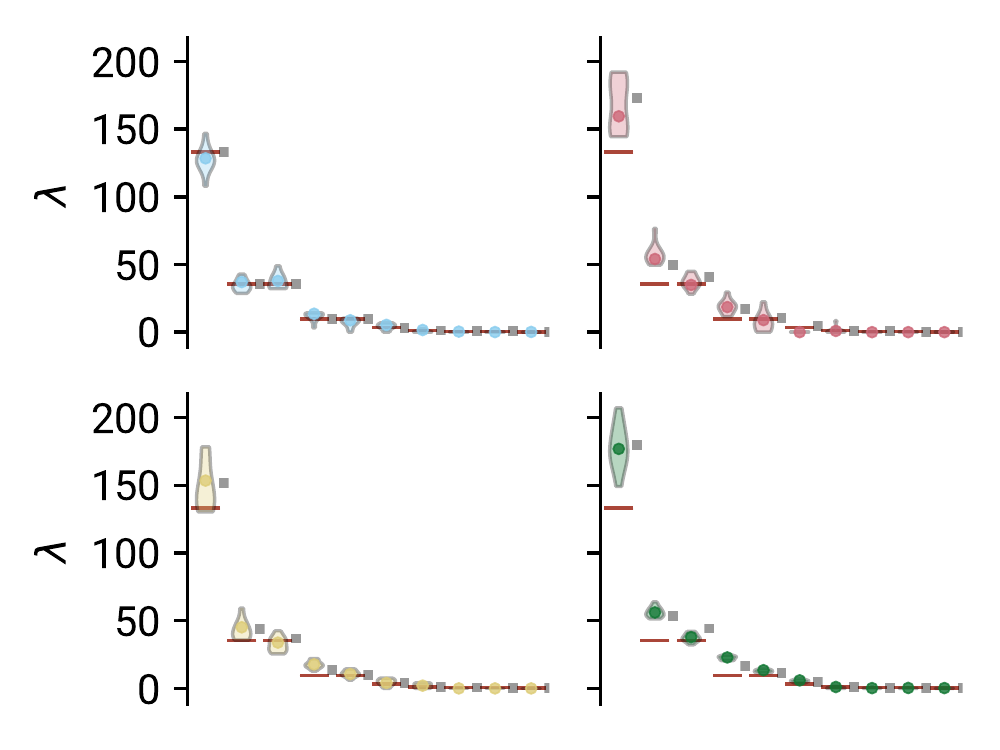}
    \caption{Comparison of eigenspectra of benzene (red lines) and predictions by each spectroscopy decoder model (violin plots); colors represent the same models as those in Figure \ref{fig:trainingerror}. The thickness of each violin plot represents the distribution of values predicted after 1,000 iterations of sampling. The filled circles represent the distribution means for each eigenvalue. Black squares represent predictions using $k$-nearest neighbors regression based on 5 neighbors.}
    \label{fig:eigenspectrumprediction}
\end{figure}

As a concrete example, Figure \ref{fig:eigenspectrumprediction} compares the ground truth eigenspectrum for benzene (\ce{C6H6})---a highly symmetric ($D_{6h}$) oblate top with $B\approx5700$\,MHz---and the corresponding model predictions when provided with the spectroscopic parameters of benzene. The violin plots represent a distribution of possible eigenvalues, given the input spectroscopic constants. Qualitatively, we see that the pure hydrocarbon model (blue) provides the closest match to the ground truth, which is contained within the uncertainty of each eigenvalue. The other models produce similar eigenspectra, with subtle differences in the magnitudes of the eigenvalues: for example, the oxygen- and nitrogen-bearing model (green) systematically predicts large leading eigenvalues, which reflects the type of molecules this model was trained with.

As a point of comparison, black squares in Figure \ref{fig:eigenspectrumprediction} predictions from a $k$-nearest neighbors algorithm as implemented in the Scikit-Learn library \cite{pedregosa_scikit-learn:_2011}, which acts as a baseline for accuracy, using the same training process as for the neural networks. Based on 5 neighbors measured by the $l^2$ distance we attain similar results to the neural network model means, although in the case of the HCO composition, overpredicts the lead eigenvalues. Although accuracy is similar between the two machine learning techniques, the $k$-nearest neighbors results are deterministic, and therefore does not provide an estimate of uncertainty. Because we are interested in performing statistical inference, it is important that uncertainty between each step is propagated appropriately.

\begin{figure}
    \centering
    \includegraphics[width=0.7\textwidth]{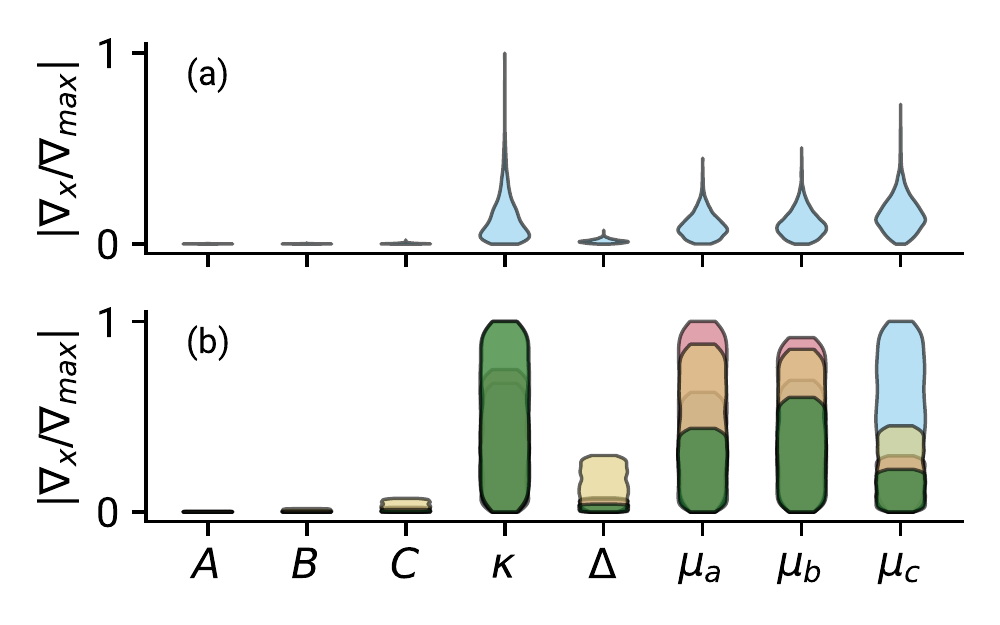}
    \caption{Violin plots of normalized, unsigned gradients computed through backpropagation of the hydrocarbon model following 3,000 iterations. The two panels represent Gaussian (a) and uniform (b) noise as inputs ($x$) to the model. Plot colors correspond to the same compositions described in Figure \ref{fig:eigenspectrumprediction}.}
    \label{fig:inputgrads}
\end{figure}

The advantage of simpler, supervised machine learning techniques is often interpretability---we show, however, that the eigenspectrum decoder can still be readily interpreted with respect to the input parameters. By design, the eigenspectrum decoder should translate input spectroscopic parameters into Coulomb matrix eigenvalues, and in through unsupervised training the model learns which parameters are more important or discriminating than others, which can be quantified via input gradients. Figure \ref{fig:inputgrads}(a) shows the distribution of gradients for each spectroscopic parameter after repeated iterations of Gaussian noise into the hydrocarbon decoder model. Here, the Gaussian noise represents some semi-structured, barely semantic information and the corresponding gradients provide an indication of how that information affects the model outputs. We see that the most informative parameters are $\kappa$, followed by $\mu_c$, $\mu_b$, and $\mu_a$, with the dipole moments on average providing more information than $\kappa$. This suggests that the model is most effectively utilized when the user has knowledge of which axes of dipole moments are non-zero, as well as the asymmetry parameter $\kappa$, and to a much lesser extent the inertial defect ($\Delta$).

Where panel (a) focuses on the hydrocarbon distribution, Figure \ref{fig:inputgrads}(b) shows the gradient distributions for each model composition using uniform noise as inputs, which should measure the true response of the network absent of any semanticity. While dipole moments and asymmetry are consistently the most defining features, each model responds quantitatively differently to each parameter, as a direct consequence of the different types of bonding and structure within each composition. Perhaps most indicative of this is the importance of $\Delta$ for nitrogen-bearing (yellow) molecules, suggesting that planarity is a much more defining characteristic and carries more variation for nitrogen molecules than for the other compositions. The most defining feature of pure hydrocarbons is the dipole moment along the $C$; structurally, this can be rationalized as an indirect measure of the number of carbon atoms along the $C$ principal axis which act as the primary source for polarization. Thus, one of the benefits of using an unsupervised, ensemble learning approach is that each model can fluidly adapt to features best suited for that particular chemical composition.

\subsection{Formula decoder}

One of the quantities that we wish to determine are possible chemical compositions: following conversion of the spectroscopic data into eigenspectra, the formula decoder model seeks to predict which and how many atoms are possible for a given eigenspectrum. Panel (b) of Figure \ref{fig:trainingerror} shows the training loss over 20 epochs: each model shows a similar loss profile, quickly converging by approximately 10 epochs. On the GV100, this corresponds to approximately 13 minutes of training time for all four models. Contrasting to the spectroscopy decoder model, the learning capacity of the formula decoder model appears to be adequate, as indicated by the closely matching training and validation curves. As the models are not overfitting, it is likely that the learning capacity could be increased, and should be considered in future architecture searches. It is important to note, however, that bias terms in the final layers were found to dominate the model outputs if unmitigated (or in our case, removed), and detrimentally affects model generalization.

\begin{figure}
    \centering
    \includegraphics[width=0.7\textwidth]{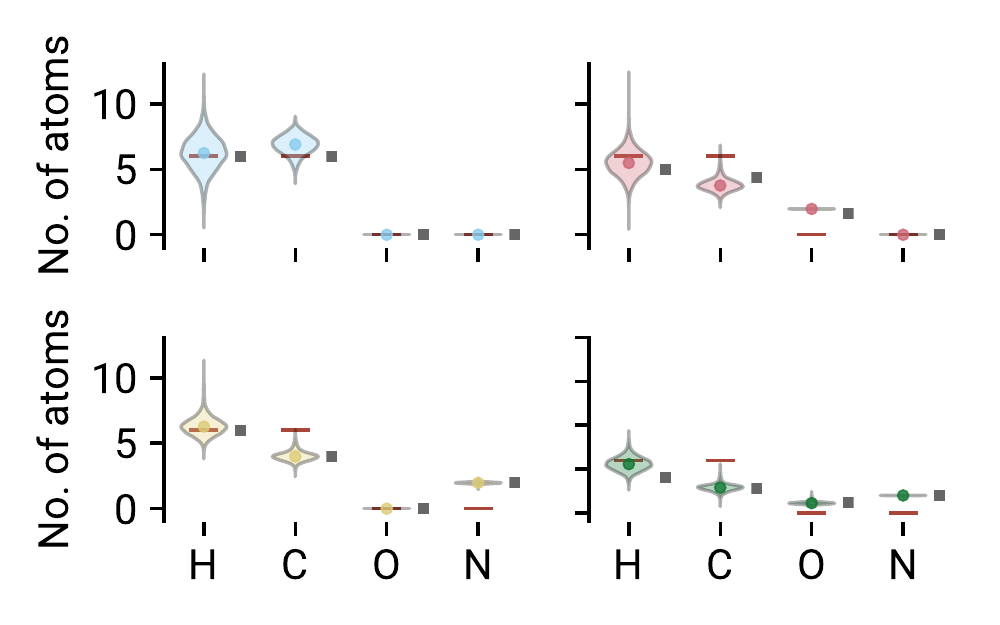}
    \caption{Predictions of chemical composition by each respective formula decoder model. Red lines represent the ground truth (\ce{C6H6}). Scatter points correspond to the expected value for each atom type after 2,000 samples. Colors refer to the same model compositions as Figure \ref{fig:trainingerror}. Black squares indicate predictions from $k$-nearest neighbors regression based on 5 neighbors.}
    \label{fig:benzeneformula}
\end{figure}

Continuing with benzene as an example, Figure \ref{fig:benzeneformula} demonstrates the performance of the combined spectroscopy and formula decoder models; each iteration involves predicting eigenspectra corresponding to the benzene constants, whereby the spectra are then passed as input into the formula decoder model. Two general trends are seen in Figure \ref{fig:benzeneformula}: first, the largest uncertainty is seen in the number of hydrogens; second, the number of heavy atoms is effectively conserved across the models -- the inclusion of oxygen and nitrogen compensates by removing carbon. Both observations are interpreted in terms of the physical properties learned by the decoder models. In the former case, hydrogen atoms are significantly lighter and therefore do not contribute much to the magnitude of rotational constants, and is appropriately reflected with a correspondingly large uncertainty. The latter trend sees that all four models conserve the effective combined mass of the molecule: there are a limited number of ways that mass can be distributed to yield the same set of rotational constants, within the constraints of atomic composition and mass. In all cases, the expected number of heavy atoms is roughly six, which matches that of benzene. These two observations not only lend confidence in the performance of the model, but more importantly that the model predictions can be rationalized with chemical intuition. While the formula decoder has marginally less accuracy than the baseline $k$-nearest neighbors (black scatter points), the ability to interpret the model uncertainty in terms of molecular structure we believe is invaluable when identifying unknown molecules.

\begin{figure}
    \centering
    \includegraphics[width=0.7\textwidth]{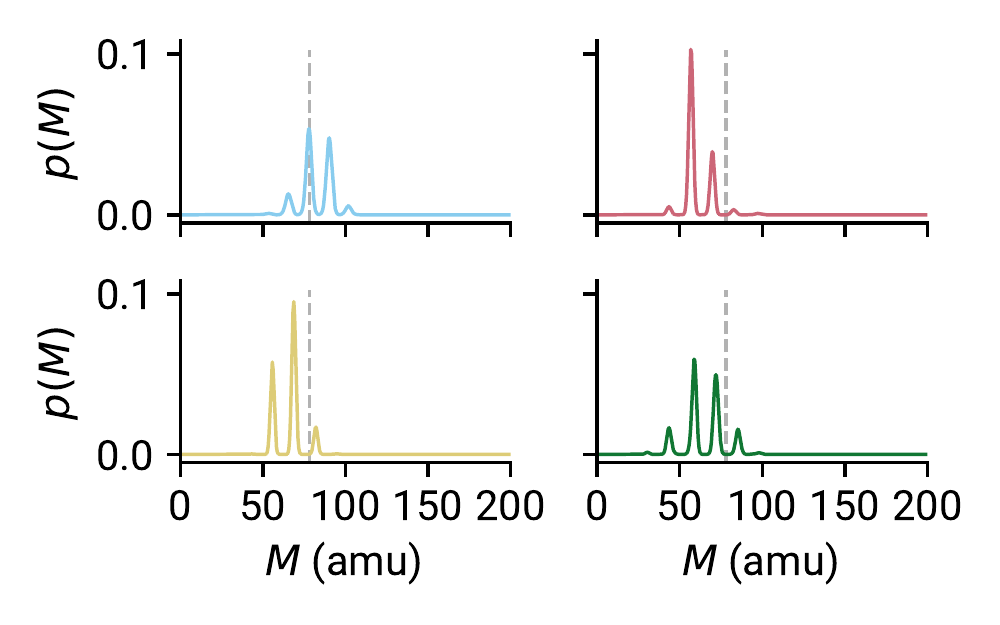}
    \caption{Simulated mass spectrum based on the predicted, quantized compositions in Figure \ref{fig:benzeneformula} by the pure hydrocarbon model. The probability distributions are obtained by Gaussian kernel estimates with a bandwidth ($\sigma$) of 1.5 mass units. The dashed line indicates the mass of benzene at 78.11\,amu.}
    \label{fig:massspec}
\end{figure}

A complementary interpretation of the predicted formulas is to generate synthetic ``mass spectra'' as shown in Figure \ref{fig:massspec}, which can be helpful when assaying unknown mixtures that have available mass resolved (e.g. spectrometry) data to compare with. The predicted compositions shown in Figure \ref{fig:benzeneformula} are quantized and used to calculate the molecular mass. Kernel density estimation is subsequently used to predict the likelihood of a given mass. The mass spectra can be interpreted in two ways: the maximum likelihood estimate (MLE) gives a point estimate of what the most likely mass that corresponds to the input spectroscopic parameters, while the distribution reflects uncertainty in the model. In the case of the pure hydrocarbon model, the MLE predicts a mass close to the ground truth (${\sim}$79.6\,amu vs. 78.11\,amu), and masses with more or fewer than six carbons are considerably less likely. The \ce{HCO} and \ce{HCNO} models yield MLE near the target, but do not reproduce the mass of benzene exactly; the eigenspectra encodes structure and composition, and the offset masses pertain to possible structures of a given composition that \textit{can} correspond with the specified parameters.

\subsection{SMILES decoder}

Figure \ref{fig:trainingerror}(c) shows the loss---as the minibatch mean KL-divergence---for the SMILES LSTM decoder model training. Each model appears to converge quickly, reaching an asymptote within several epochs, and required ${\sim}$50 minutes of computation on a GV100. The ultimate training and validation accuracy of each model is quite exceptional: the worst performer, the pure hydrocarbon model, yields a KL-divergence averaged across the entire sequence of ${\sim}$0.15, close to the minimum possible value of zero.\cite{kullback_information_1978} Thus, according to this loss metric the model is able to reproduce the long sequences of encodings accurately without under or overfitting.

Where the composition is a helpful quantity, the ultimate goal is to determine possible structures that can be assigned to the spectroscopic parameters. There are a variety of formats that this information can be conveyed, for example as simple Cartesian structures, or reconstructing the Coulomb matrix based on the predicted eigenvalues, or as string identifiers such as SMILES \cite{oboyle_towards_2012} and InChI \cite{heller_inchi_2015}. The string representations are particularly attractive encodings as they are machine and human parsable, and in certain forms (for example, canonical SMILES) and can discriminate enantiomers. For our purposes, we chose canonical SMILES as a target due to its simplicity: in contrast, the syntax for InChI is extremely specific, and unlikely to be fully reproduced with the limited amount of experimental information. SMILES strings, even when incomplete, can be used to infer likely functional groups and using programs such as OpenBabel, can be used to generate initial guess Cartesian structures for subsequent optimization with electronic structure methods. Because of its wide use in cheminformatics for drug discovery and reaction screening, there have been multiple applications of deep learning that utilize SMILES; recurrent approaches such as LSTM\cite{hochreiter_long_1997} and GRU architectures\cite{cho_learning_2014} are best suited for sequence-to-sequence translation, whereby one SMILES string is used to predict another.\cite{arus-pous_randomized_2019}

\begin{figure}
    \centering
    \includegraphics[width=0.8\textwidth]{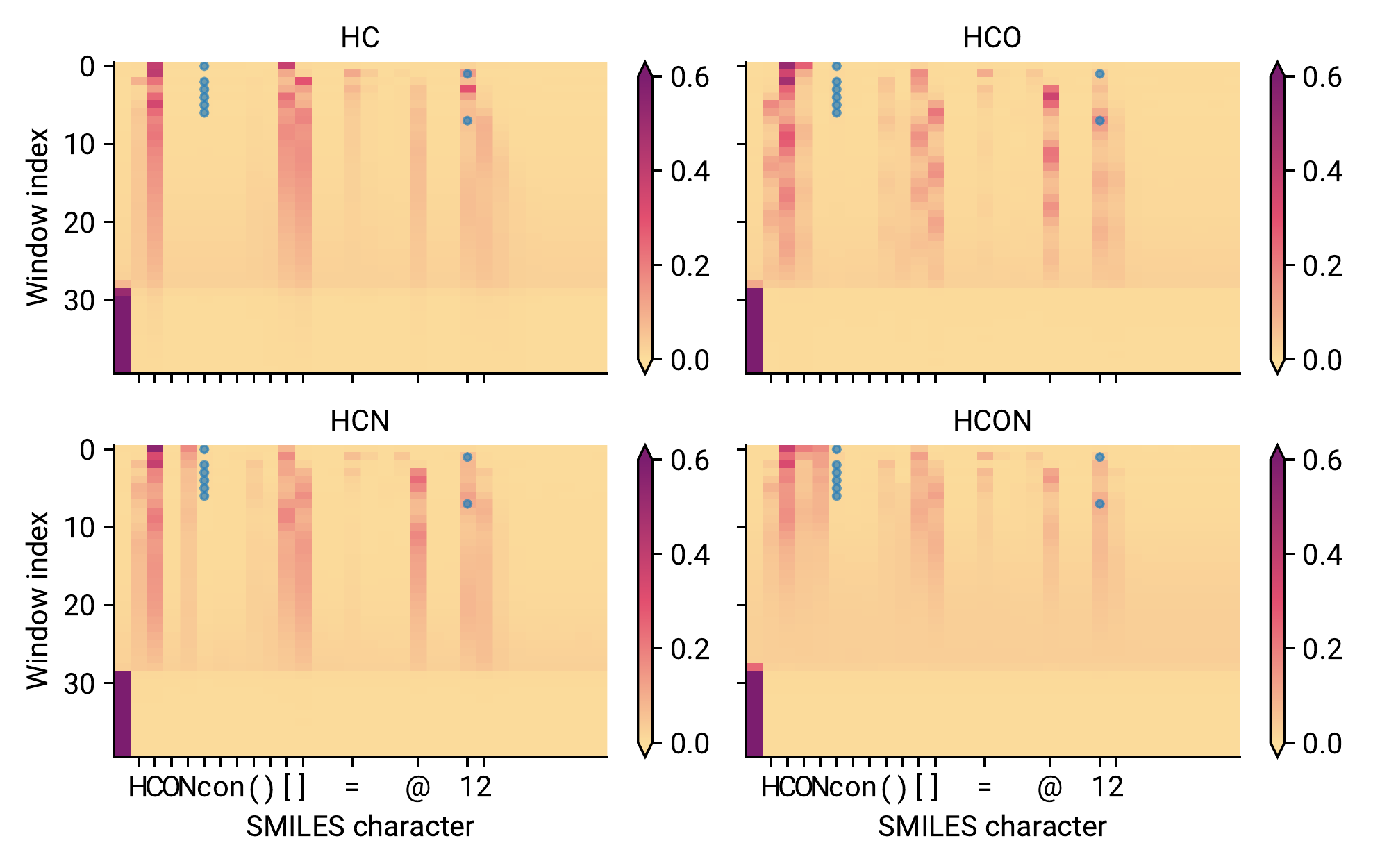}
    \caption{Heatmap of SMILES character probabilities predicted using spectroscopic parameters of benzene averaged over 2,000 samples, truncated for the first 40 sequences. Each quadrant represents the same model compositions as previous figures, as indicated in the quadrant titles. The abscissa and ordinate correspond to the SMILES character encoding index and the sequence window index respectively; the first SMILES encoding corresponds to an empty character. Progressively darker colors correspond to higher probabilities for a given symbol. Blue dots indicate the ground truth encoding for benzene; $\mathrm{c1ccccc1}$.}
    \label{fig:benzenesmiles}
\end{figure}

In our application, we convert sequences of eigenspectra into SMILES characters with LSTM units: each window of eigenvalues are used to predict the likelihood of each symbol within our SMILES corpus, and due to the recurrent nature of the LSTM architecture, the hidden outputs of each window are used to predict the likelihoods of following windows. The rationale is to recover nuances of SMILES syntax; for example a closing bracket may appear several or many characters after an opening which indicate side branching in a chain. Similarly, a closing bracket should not appear prior to an opening one. Figure \ref{fig:benzenesmiles} visualizes the outputs from the SMILES decoder model based on benzene parameters, truncated to the first 40 sequence windows: the heatmap represents the averaged likelihood of a character within our corpus (abscissa) for a given sequence window (ordinate). These averages are useful for illustrating what semantics are learned by the LSTM model. We see that in all four compositions, the string terminates at a sequence length of approximately 30 characters, whereby the likelihood maximizes on the whitespace character. This indicates that the model learns an appropriate length and complexity of a SMILES string from its eigenspectrum. Another general observation is that the most likely character in early windows regardless of model composition is aliphatic carbon (C) -- because all molecules contain mostly carbon, associating a high likelihood with carbon becomes inevitable. Later in the sequence, other characters become more likely, including other elements and bonding specification. One of the more important features is the ordering of parentheses appears to be successfully learned by the model, whereby the likelihood of a closing bracket is zero initially until an opening bracket has non-zero probability of appearance -- this is the intended consequence of using a LSTM model.

The major obvious shortcoming of our model, however, is that it fails to reproduce the SMILES code of benzene ($\mathrm{c1ccccc1}$, indicated by the blue scatter points). Upon inspection of our training set, it appears that only ${\sim}$3200 molecules contain the aromatic carbon symbol, and is therefore significantly underrepresented and unsurprising that little to no likelihood is predicted by the models. On a more general note, the encoding for benzene is highly unique because of its molecular symmetry ($D_{6h}$), thus it is unlikely a generalized LSTM model can successfully reproduce the specificity required for molecules like benzene; in other words, there is ``no free lunch''.\cite{wolpert_no_1997,goodfellow_deep_2016} This example highlights the limitation of our SMILES decoder model, where highly symmetric---and typically small---molecules are poorly reproduced due to their high specificity and symmetry in favor of large asymmetric species. We contend, however, that these molecules are the most difficult to identify and in need of an inferential approach contrasting the smaller molecules that can be more readily deduced via combinatorial searches.

Following calculation of the character likelihoods, we employed a beam search algorithm for decoding the sequences into SMILES strings. This is performed by starting with $n$ of the most likely characters at the beginning of the sequence, and finding characters along the sequence that maximizes the conditional likelihood. In spite of this, we find many of the resulting strings to be invalid SMILES syntax, particularly with the placement and ordering of parentheses, and often chemically vague. Another issue we observe is the coherence time of the sequences: in many samples, we see that the character likelihoods decay gradually into approximately uniform likelihoods, as the eigenvalues are effectively zero and the LSTM model fails to produce any information. This criterion is used during the beam search, where the sequence likelihood is compared against a uniform distribution using the KL-divergence: as it approaches zero, sampling is terminated early to prevent oversampling from uninformative sequences. 

As we see in Table \ref{tab:smiles}, the highest conditional likelihood strings are unfortunately chemically and structurally uninformative. The most striking issue is aliphatic carbon is significantly oversampled, most likely due to the fact that the dataset contains organic molecules---and with little information available---the most likely character within a SMILES sequence will be carbon. Another problem is the length of the sequences: even with early termination, the sequences produced are far too long to match the rotational constants of benzene. In the models containing oxygen and nitrogen, we see that these elements are incorporated into the sequence, albeit extremely unlikely; for example, CCCCOOO in the HCON model. To improve this approach, future attempts should consider changing different aspects of the problem: for example, the eigenspectrum is not necessarily an optimal feature representation to decode into SMILES strings, and could be advanced by projection onto a more informative space (i.e. principal components) or other machine readable representations.\cite{wu_moleculenet_2018,huang_communication_2016} The neural network architecture could also be substantially improved upon, for example by using transformer architectures.\cite{vaswani_attention_2017} Finally, the information content of SMILES could be encoded in different ways, such as lossless compression\cite{baldi_lossless_2007}. While we used a one-hot approach successfully demonstrated by other groups\cite{hirohara_convolutional_2018, arus-pous_randomized_2019,winter_learning_2019} for direct SMILES-to-SMILES translation, it is likely that the uncertainty is too high in our application for unique and informative mapping. Various forms of SMILES compression, such as DeepSMILES\cite{arus-pous_randomized_2019}, would greatly simplify the encoding complexity and decrease the machine learning requirements---an avenue for future exploration.

\begin{center}
\begin{table}[]
    \begin{tabular}{l l l l}
        HC & HCO & HCN & HCON \\
        \toprule
         CCCCCCCCCCCC & CCCCCCCCCC & cCCCCCCCCCCCCC & OCCCCOO \\
         CCCCCCCCC & OCCCCCCCCC & CCCCCCCCCCCCCCC & NCCCCOO \\
         CCCCCCCC & OCCCCCCCCCCC & NCCCCCCCCCCCCCCCC & CCCCOOO \\
         CCCCCCCCCCCCCC & OCCCCCCCCCCCC & nCCCCCCCCCCCCCCC & OCCCCONC\\
    \end{tabular}
    \caption{Four SMILES strings with the highest conditional likelihoods based on 2,000 iterations of sampling, decoded with the beam search algorithm. Predictions are based on the spectroscopic constants of benzene.}
    \label{tab:smiles}
\end{table}
\end{center}

\subsection{Functional group classification}

As the SMILES LSTM decoder---in its current state---was unable to produce useful information for molecular identification, we investigated the possibility of simpler, yet indicative sources of information. Combining 1D convolution and linear layers, we built a model that uses the eigenspectra and the spectroscopic parameters to perform multilabel classification, which predicts the likelihood of selected functional groups being present. This is premised by the fact that the parameters, in particular the dipole moment vectors, contains some information about functional groups which are the primary drivers for polarization in a molecule. Combined with the eigenspectra, there should be sufficient information to reliably distinguish between similar yet different functional groups (for example, alcohol group within carboxylic acids and primary alcohols).

Figure \ref{fig:trainingerror}(d) shows the training and validation binary cross-entropy profiles over 40 training epochs. On a GV100, model training took approximately ${\sim}$11 minutes to complete. Once again, both training and validation losses are nearly identical, indicating that the models are neither over or underfit. The hydrocarbon model demonstrates exceptionally low loss, which is ascribed to low chemical complexity, as there are few functional groups that are possible. While HCO and HCON models show the largest loss values, considering the full breadth of functional groups are possible (15 labels for the former, 23 for the latter), we believe each model is performing within the full capacity of the architecture.

In multilabel classification, the binary cross-entropy alone is not informative of the model performance. Using $k$-nearest neighbors as an unsupervised baseline classifier, we performed approximately the same multilabel classification task as with the neural network model. For comparison, we use the $F_1$-score which is the harmonic mean of the precision and recall scores; the former measures the number of times the correct label is predicted out of the total number of samples, whereas the latter represents the number of times the correct label is predicted, divided by the number of examples of that label. An $F_1$-score of unity represents the case where every label was correctly predicted at every possible instance, and not simply from random chance.

\begin{figure}
    \centering
    \includegraphics[width=0.5\textwidth]{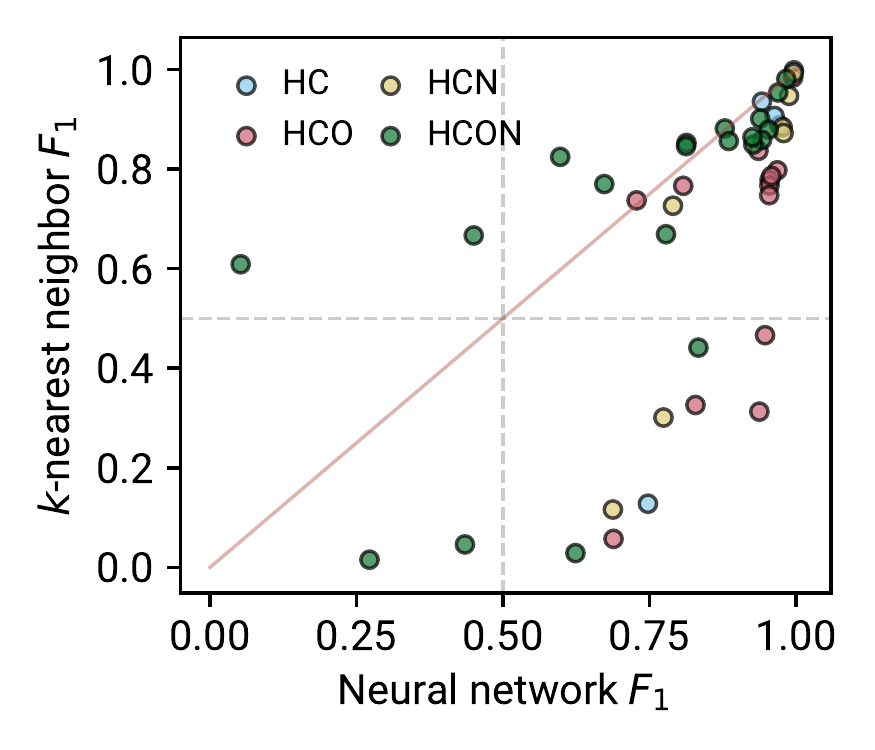}
    \caption{Comparison of validation $F_1$ scores from the neural network approach (abscissa) and a $k$-nearest neighbors classifier (ordinate). Each scatter point corresponds to a functional group encoding, where colors represent the compositions used to train each respective model. The solid red trace indicates where both models perform equally well.}
    \label{fig:baselineclassifier}
\end{figure}

Figure \ref{fig:baselineclassifier} compares the $F_1$ scores calculated by the two approaches, for each functional group within a composition. In many cases, both classifier models show excellent performance (top right quadrant) where $F_1$ is close to unity. There are, however, several functional groups within the HCON composition that are not predicted well in either classification model which are vinyl groups, carbonyls, and alcohol groups. Table \ref{tab:prf} shows the worst performing functional groups with respect to $F_1$ scores: we see that the neural network has consistently higher recall scores than precision, indicating that these functional groups are subject to false positives. In comparison, the $k$-nearest neighbors approach results in a higher precision, albeit with significantly lower recall scores and more reluctant to predict these groups. It is likely that the features are weakly discriminative with respect to these oxygen functional groups, compared to their nitrogen counterparts correspond to much higher $F_1$ scores. This is true for the allene functional group, which suffers from consistently lower predictability across all four compositions by both classification models.

\begin{table}[]
    \centering
    \begin{tabular}{llrrrrrr}
    \toprule
    ~ & ~ & \multicolumn{3}{l}{Neural network} & \multicolumn{3}{l}{$k$-nearest neighbors} \\
    \midrule
    Model & Functional &  Precision &  Recall &  $F_1$ &  Precision &  Recall &  $F_1$ \\
    \midrule
     HCON &                   Allene &         0.08 &      0.21 &        0.05 &           0.78 &        0.50 &          0.61 \\
     HCON &                    Vinyl &         0.38 &      0.63 &        0.27 &           0.46 &        0.01 &          0.02 \\
     HCON &                    Ether &         0.49 &      0.56 &        0.43 &           0.43 &        0.02 &          0.05 \\
     HCON &               Amino acid &         0.52 &      0.63 &        0.45 &           0.84 &        0.55 &          0.67 \\
     HCON &                   Alkyne &         0.67 &      0.76 &        0.60 &           0.86 &        0.80 &          0.82 \\
     HCON &  Carboxylic acid alcohol &         0.65 &      0.69 &        0.62 &           1.00 &        0.01 &          0.03 \\
     HCON &                 Peroxide &         0.73 &      0.80 &        0.67 &           0.78 &        0.76 &          0.77 \\
      HCN &                    Vinyl &         0.70 &      0.72 &        0.69 &           0.67 &        0.06 &          0.12 \\
      HCO &                   Phenol &         0.72 &      0.76 &        0.69 &           0.50 &        0.03 &          0.06 \\
      HCO &                   Allene &         0.74 &      0.75 &        0.73 &           0.78 &        0.70 &          0.74 \\
    \bottomrule
    \end{tabular}

    \caption{Lowest ten $F_1$ scores for the neural network approach, comparing the precision and recall scores for both classification models.}
    \label{tab:prf}
\end{table}

The key observation from Figure \ref{fig:baselineclassifier} is the superior performance by the neural network classifier. We can conclude that the neural network approach is better suited for molecular identification, with three distinct advantages over the baseline model: (1) improved precision and recall in all except one functional group, (2) uncertainty quantification through dropouts, and (3) portability and scalability. The last advantage is particularly important towards real-time inference; the $k$-nearest neighbor classifier needs to traverse the full training set (9\,GB of data) for inference, thus scaling poorly with the data set size and limiting portability. On the other hand, the neural network classifier is significantly compressed (${\sim}2.6$\,MB on disk), and can be used in distributed systems and GPUs.

In terms of the performance of the neural network, Table \ref{tab:bestclassifier} shows the top 15 performing $F_1$, precision, and recall scores, with their respective composition and functional group. These metrics show that, in the best case scenarios, the classifier is able to predict the presence of a functional group to ${\sim}85\%$ precision, simply from a set of spectroscopic parameters. Most importantly, it is difficult to establish a human judgement baseline, as it is highly unlikely that an expert is able to derive such information simply from inspecting rotational constants and dipole moments. This is extended to the vast majority of the functional groups included in our study: Table \ref{tab:prf} shows the worse performers with respect to $F_1$ scores, such that $>75\%$ of the predictors are accurate to 70\%.

\begin{table}[]
    \centering
    \begin{tabular}{llrrr}
    \toprule
    Model &         Functional &  Precision &  Recall &  F1-score \\
    \midrule
      HCN &    Aromatic carbon &       0.96 &    0.93 &      0.99 \\
      HCN &             Alkyne &       0.84 &    0.74 &      0.98 \\
      HCN &            Nitrile &       0.98 &    0.97 &      0.98 \\
       HC &             Allene &       0.84 &    0.75 &      0.97 \\
     HCON &    Aromatic carbon &       0.96 &    0.95 &      0.97 \\
      HCO &           Peroxide &       0.96 &    0.95 &      0.97 \\
       HC &             Alkyne &       0.92 &    0.88 &      0.96 \\
      HCO &           Aldehyde &       0.90 &    0.84 &      0.96 \\
      HCO &           Carbonyl &       0.95 &    0.94 &      0.95 \\
      HCO &    Carbonyl-carbon &       0.94 &    0.93 &      0.95 \\
      HCO &             Ketone &       0.88 &    0.82 &      0.95 \\
     HCON &           Carbonyl &       0.91 &    0.86 &      0.95 \\
      HCO &            Alcohol &       0.92 &    0.90 &      0.95 \\
     HCON &  Carbonyl-nitrogen &       0.86 &    0.79 &      0.94 \\
       HC &    Aromatic carbon &       0.92 &    0.90 &      0.94 \\
    \bottomrule
    \end{tabular}
    \caption{Top 15 performing functional groups and their associated statistics for each neural network classifier composition, based on the validation dataset.}
    \label{tab:bestclassifier}
\end{table}

\begin{figure}
    \centering
    \includegraphics[width=0.7\textwidth]{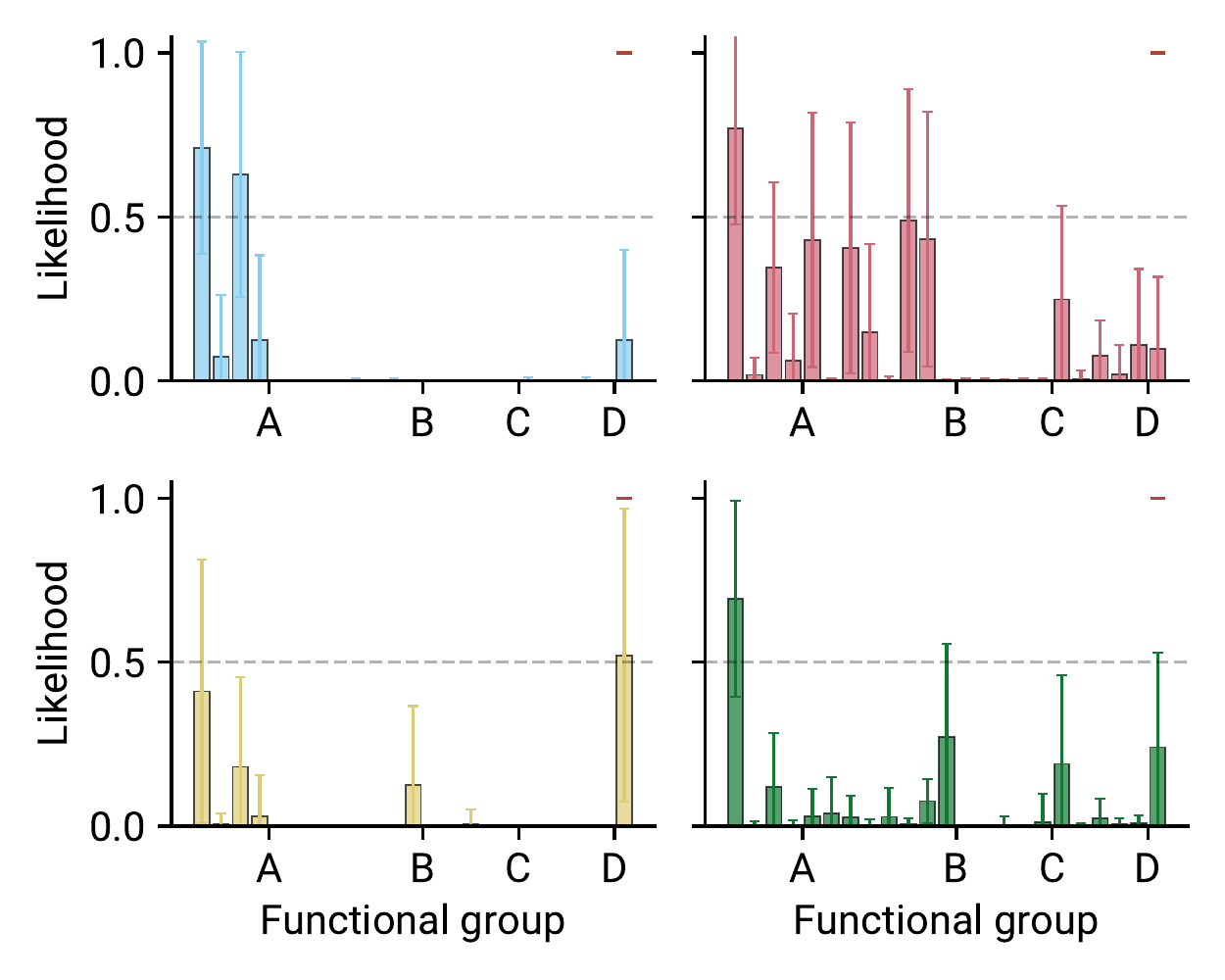}
    \caption{Predicted mean likelihoods of each functional group by each respective model composition. Error bars represent 1$\sigma$ in model uncertainty. The dotted marks an arbitrary cut-off of 50\% likelihood. The absicssa labels represent types of functional groups to the left of the label: (A) carbon saturation, (B) carbonyls, (C) nitrile/nitro, (D) alcohol/acid. The last group corresponds to aromatic carbon; see Table \ref{tab:grouplabels}.}
    \label{fig:functionalgroup}
\end{figure}

Figure \ref{fig:functionalgroup} continues to use benzene as a demonstration, where each quadrant represents shows the predicted functional groups for a given composition. Given the large number of labels, we defer the reader to Table \ref{tab:grouplabels} for a list of the labels within each label group. The outputs of this classifier predicts, as shown by each bar, the likelihood of a particular functional group being present in the molecule given the Coulomb matrix eigenvalues and spectroscopic parameters. A full ordered list of the functional groups is given in Table S1. In the case of the pure hydrocarbon model, the most likely groups predicted are aliphatic carbon and vinyl groups, and with much lower probability aromatic carbon followed by alkyne, with the least likely an allene group. Although the model incorrectly ascribes lower probability to the correct (aromatic carbon) label, it does infer a high likelihood of unsaturation via the vinyl group. On the other hand, the nitrogen (yellow) and mixed (green) models predict a high likelihood of aromaticity. Interestingly, the oxygen-bearing (red) model predicts a large likelihood for many functional groups, particularly those pertaining to carbonyls (between A and B).

\begin{table}[]
    \centering
    \begin{tabular}{l l}
        Label group & Functional group  \\
        \toprule
        A & Aliphatic \\
        ~ & Allene \\
        ~ & Vinyl \\
        ~ & Alkyne \\
        B & Carbonyl \\
        ~ & Carbonyl-nitrogen \\
        ~ & Carbonyl-carbon \\
        ~ & Aldehyde \\
        ~ & Amide \\
        ~ & Ketone \\
        ~ & Ether \\
        C & Amine \\
        ~ & Amino acid \\
        ~ & Nitrate \\
        ~ & Nitro \\
        D & Alcohol \\
        ~ & Carboxylic acid\\
        ~ & Enol \\
        ~ & Phenol \\
        ~ & Peroxide \\
        ~ & Aromatic carbon\\
        \bottomrule
    \end{tabular}
    \caption{Ordering of the functional group labels.}
    \label{tab:grouplabels}
\end{table}

The probabilistic approach we have adopted here allows a user to consider not only the probability of a functional group, but also the model confidence. Furthermore, because the labeling is generated by matching SMARTS substructures, one can easily create arbitrarily specific functional group classification schemes; in the current implementation we chose to use quite general SMARTS coding to maximize coverage, however this could be tuned to produce highly specific labels (for example heteroatomic ring structures). In the proceeding sections, we will discuss how predictions from each of the models can be combined to infer the identity of an unknown molecule, or at least suggest tests to be conducted.

\subsection{Example applications}

\begin{figure}
    \centering
    \includegraphics[width=\textwidth]{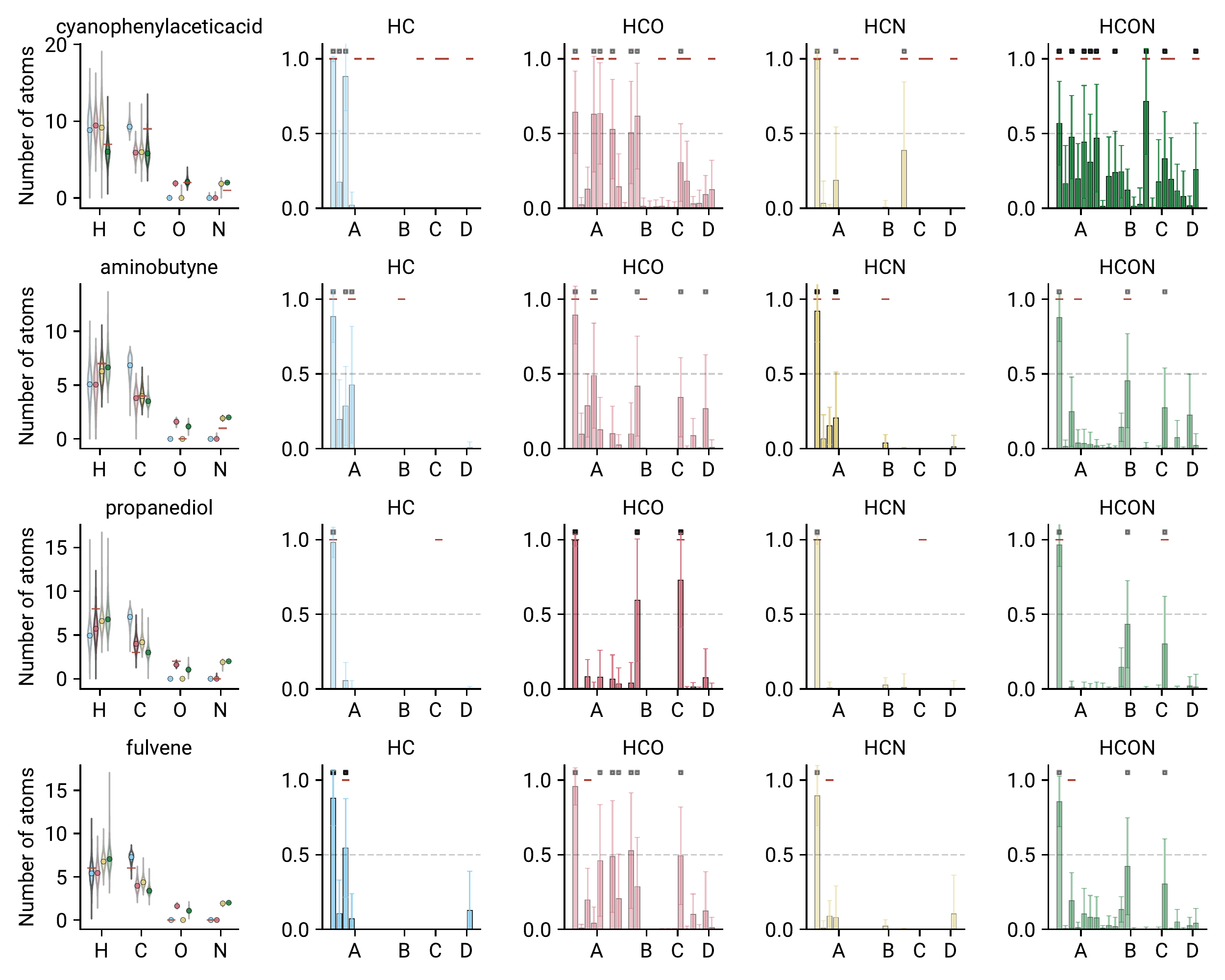}
    \caption{Mosaic of the predicted distributions of molecular composition (left-most panels) and likelihoods of functional groups for four selected species. In each panel, red lines indicate the ground truth. In the functional group predictions, bars represent the mean prediction with $1\sigma$ uncertainties shown in the error bars, and black squares indicate functional groups with likelihoods greater than 0.5. Darker shading corresponds to the correct model composition. The absicssa labels represent types of functional groups to the left of the label: (A) carbon saturation, (B) carbonyls, (C) nitrile/nitro, (D) alcohol. The last group corresponds to aromatic carbon.}
    \label{fig:demonstration}
\end{figure}

In this section, we will apply the formula and functional group decoder models to four known molecules in order to demonstrate our anticipated workflow/thought process. Generally speaking, the formula decoder sets the boundaries for viable compositions, and combined with the predicted functional groups should significantly limit the search space. We note that these examples were not chosen based on their performance, rather as a way to highlight the strengths and weaknesses of the models outlined in this work, and how the predictions from each model can be combined to piece together information about an unknown molecule. Figure \ref{fig:demonstration} shows the predictions on four different species by the formula decoder and functional group classifier by each model composition. Inference with all four models was performed on a Nvidia V100 GPU with 5,000 samples per molecule, at approximately $4-6$ seconds per molecule. 

Starting with cyanophenylacetic acid (\ce{C9H7NO2}), an aromatic molecule with nitrile and carboxylic groups, we see that the number of atoms predicted by the (HCON, green) formula decoder is fairly accurate, although the number of nitrogens is overpredicted and not captured by the model uncertainty, thereby showing the model remains overconfident in spite of data augmentation. The corresponding functional group classifier (HCON) correctly predicts six out of seven groups---missing only the carboxylic acid group which, based on the $F_1$ scores in Table \ref{tab:prf}, is one of the poorly captured groups by the HCON classifier. Additionally, there are three other false positive predictions which are a vinyl group, a nitrogen atom in the $\alpha$ of a carbonyl, and an amide group. This result reinforces the fact that the current model implementation is more likely to generate false positives (i.e. low precision scores).

The next example, aminobutyne, is a typical unsaturated nitrogen-bearing molecule. In this case, the formula decoder (HCN) overestimates the number of nitrogens, although captures the number of hydrogens and carbons perfectly. The functional group classifier correctly predicts the presence of aliphatic carbon, an alkyne group, although ascribes a low likelihood for an amine group. Unfortunately, this is an example of which the functional group classifier is misleading in its prediction: from this, we recommend that these classifier models be used to guide what groups \textit{may be present}, rather than completely ruling out groups entirely.

Propanediol is an example where both the formula decoder and functional group classifier provides accurate predictions. In the latter, both the aliphatic content and alcohol functional groups are correctly predicted, along with a false positive ether group. We see here that each model composition recognizes the highly saturated nature of the input species---predicting low likelihoods for unsaturated groups (e.g. vinyl, alkene, etc.) and dominated solely by aliphatic carbon. This example highlights how predictions from each composition can jointly inform the user what common functional groups are present.

Finally, fulvene is an isomer of benzene (\ce{C6H6}). The formula decoder once again captures the number of atoms well, although the expected number of carbons is slightly higher than the actual. In contrast to the benzene example (Figure \ref{fig:functionalgroup}), none of the models predict a significant likelihood of aromatic carbon being present, and instead see a high likelihood of unsaturated alkenes (compared to the propanediol result). 

Based on the four examples, we can conclude three aspects that will guide interpretation of these models. First, the formula decoders are likely to underestimate the total number of non-hydrogen atoms, although constrain the possible formula to within an atom. When considering the possible range of formulae, it is therefore recommended to test the mean formula first, followed by modifications to the heavy atoms (in particular oxygen and nitrogen) according to the uncertainty of each atom. Because the uncertainties are often underestimated---as seen in the nitrogen and oxygen predictions---we would recommend extending the sampling of the number of atoms by $\pm1$ beyond the limits of the uncertainties. Second, the functional group classifiers appear are more likely to produce false positives than false negatives, based on Figure \ref{fig:demonstration} as well as some of the precision and recall scores shown in Table \ref{tab:prf}. Thus when testing functional groups, we recommend prioritizing the high likelihood functional groups that fall under the composition constraints, and systematically ruling each group out over the course of the identification process. These can be confirmed experimentally often by rare isotopic substitution, for example shifting alcohol groups with deuterium. Third, when the composition is unclear it is important to consider predictions from all four compositions, in particular functional groups that are common to other compositions. The most decisive trend are saturated/unsaturated species: in Figure \ref{fig:demonstration}, unsaturated species are predicted to have unsaturated content regardless of the model composition, while saturated species generally result in no unsaturated groups at all (as in the case of propanediol).

\subsection{Model considerations and limitations}

In the examples provided so far, the models are provided a complete set of spectroscopic parameters with absolute precision. In real applications, this may not always be the case; for example when combinations of parameters are being used to fit effective Hamiltonians (e.g. $B+C$ for a prolate symmetric top), or when the dipole moments are not known. The advantage of our probabilistic approach is the ability to perform inference even under these circumstances, because each model provides an estimate of the conditional likelihood $p(y \vert x)$, each parameter within $x$ can simply be varied proportional to its uncertainty and with spectroscopic intuition. To pose an example, we discuss a situation commonly encountered in our laboratory\cite{lee_study_2019}: a prolate symmetric top is fit with $B+C$ without immediately obvious $K$ structure, and only $a$-type transitions are measured thus leaving $A$ poorly constrained, and $\mu_b$ and $\mu_c$ unknown. The parameters can be repeatedly perturbed with Gaussian noise weighted by the parameter uncertainty in a bootstrap fashion. Due to the probabilistic nature of our models, the uncertainties propagate from the input values, through the eigenspectrum and to the predicted quantities; each pass is equivalent to computing the conditional likelihood of a formula or functional group with respect to $\lambda$ and $A(BC)$.

A detail that arose during the training of these models was the importance of a balanced dataset, which was particularly apparent in the functional group classifier. Despite our efforts to balance the dataset prior to training, the models produced are still susceptible to biases that are created inadvertently by unbiased sampling as we have done. For example, in our tests on smaller molecules which are underrepresented with respect to larger species, simply due to the number of possible isomers for the latter case. Future attempts of these models will need to be highly mindful of these subtleties at the possible expense of selection bias.

One of the significant drawbacks of our approach towards probabilistic neural networks is the overuse of dropout layers: although they are necessary for the probabilistic aspect of our solution, it is likely that they over-regularize parameter learning and consequently decrease the full learning capacity of each model. In principle, one could use a reduced dropout probability during training---as long as there is no overfitting---and use a larger dropout rate for inference. There are also methods to calibrate uncertainties by empirical scaling\cite{kuleshov_accurate_2018} which could rectify model uncertainties, thereby mitigating ``over-dropping''. Regardless, dropout acts only as an approximation to Bayesian sampling, and for a truly probabilistic approach inference must be performed by sampling from posterior distributions of learned parameters. A major difficulty in implementing true Bayesian deep learning models is the computational cost associated with training and inference; every forward pass must involve sampling from hundreds to thousands of parameter distributions that replace scalar values, and every backward pass must compute, propagate, and update gradient information to the same number of parameters.\cite{goodfellow_deep_2016} Bayesian networks are an active area of study, and there are attractive solutions being developed including probabilistic backpropagation, \cite{hernandez-lobato_probabilistic_2015,blundell_weight_2015} bootstrap methods\cite{rohekar_bayesian_2018}, and approximate\cite{hinton_keeping_1993} and variational\cite{graves_practical_2011} inference. The ability to move to a Bayesian model would remove the need for an ensemble approach, which would significantly improve model ease of interpretation. Here, an ensemble is required due to the difficulty for single network models to generalize and be predictive with a large variety of input parameters whereas Bayesian models are resistant to overfitting.

Overall, the proof-of-concept models we have shown here highlights the viability for probabilistic deep learning models in molecule identification with rotational spectroscopy. While there is room for improvement, the approaches we have described provide a promising framework for performing inference on unknown molecules: we can reliably constrain the possible range of compositions and functional groups present \textit{simply from a set of eight spectroscopic parameters}. These constraints---in conjunction with user expertise---can be used to guide systematic electronic structure calculations to provide possible candidates for identification. The framework we have described here has significant implications for using rotational spectroscopy in complex mixture analysis. In addition to providing a systematic method for identification, each decoder model connects rotational spectroscopy with other analytical techniques: through the formula decoder, we are able to predict mass spectra, and with the functional group classifier, we unlock an aspect of chemistry that was not previously accessible solely with rotational spectroscopy, as functional groups are typically determined using infrared techniques. By further developing this methodology, we believe this will solidify rotational spectroscopy as a universal analytical tool.

\section{Conclusions}

In this work, we demonstrated a series of proof-of-concept probabilistic deep learning models that aim to assist with molecular carrier inference. The architectures we have described are relatively simple and light-weight neural network models. In our demonstrations, we show that the approximate formula can be determined and what functional groups are likely to be present from spectroscopic data routinely available from broadband chirped-pulse experiments---the spectroscopy decoder, formula decoder, and functional group classifier can be collectively used to infer discriminating factors about the unknown molecule, which should systematically lead to its identification. Although the SMILES LSTM decoder could not generate sufficiently coherent SMILES sequences, our results show that the models proposed here are able to learn some of the semantics although it is unclear whether there is sufficient specific information contained within the eigenvalues to perform a direct translation to canonical SMILES. Instead, it may be worthwhile to consider compressed SMILES encodings, or other representations of molecular structure.

The models we have presented as part of this work are computationally scalable, and with appropriate algorithmic optimizations could provide a step towards near real-time unknown molecule inference. Furthermore, the probabilistic framework we have detailed can readily accommodate for ``real'' situations: particularly those where certain spectroscopic parameters are highly uncertain, by using bootstrapped parameters during inference. We anticipate that these models will be highly invaluable in future broadband assays of unknown, complex mixtures using rotational spectroscopy.

\begin{acknowledgement}

The authors acknowledge financial support from NSF grants AST-1615847 and AST-1908576, and NASA grants NNX13AE59G and 80NSSC18K0396, and computing resources from the Smithsonian Institution High Performance Cluster (SI/HPC, ``Hydra'').

\end{acknowledgement}

\begin{suppinfo}

Dataset used for the model training can be made available upon request. A table containing information regarding the functional group labels is included as Supporting Information. The Python code used to train, test, and perform inference can be found on https://github.com/laserkelvin/rotconml.

\end{suppinfo}

\bibliography{references}

\end{document}